\documentclass{birkjour}

\usepackage[T2A]{fontenc}
\usepackage[english]{babel}
\usepackage{latexsym}
\usepackage{amssymb}
\usepackage{amsmath}

 \newtheorem{thm}{Theorem}[section]
\def\proof{\medbreak\noindent{\bf Proof}}
\def\theorem #1. #2\par{\medbreak
  \noindent{\tt {\bf Theorem #1.}\enspace}{\sl#2\par}%
  \ifdim\lastskip<\medskipamount \removelastskip\penalty55\medskip\fi}
\def\lemma #1. #2\par{\medbreak
  \noindent{\tt {\bf Lemma #1.}\enspace}{\sl#2\par}%
  \ifdim\lastskip<\medskipamount \removelastskip\penalty55\medskip\fi}

\def\{{\lbrace}
\def\}{\rbrace}

\def\Wcl{W{\mathbb C}\!\ell}
\def\wcl{w{\mathbb C}\!\ell}

\def\cl{{\mathbb C}\!\ell}

\def\R{{\Bbb R}}
\def\C{{\Bbb C}}

\def\L{{\cal L}}

\def\O{{\rm O}}

\def\be{\begin{equation}}
\def\ee{\end{equation}}

\newcommand{\A}{{\mathbb A}}
\newcommand{\bbE}{{\mathbb E}}
\newcommand{\bbI}{{\mathbb I}}
\newcommand{\bbJ}{{\mathbb J}}
\newcommand{\bbK}{{\mathbb K}}
\newcommand{\st}{\stackrel}
\newcommand{\vsp}{{\vrule width0pt height15pt}}

\begin{document}

\title[Quaternion typification of Clifford algebra elements]{Quaternion typification\\ of Clifford algebra elements}

\author[D.~S.~Shirokov]{D.~S.~Shirokov}
\address{Steklov Mathematical Institute\\
Gubkin St.8, 119991 Moscow, Russia}
\email{shirokov@mi.ras.ru}
\begin{abstract}
We present a new classification of Clifford algebra elements. Our classification is based on the notion of quaternion type. Using this classification we develop a method for analyzing commutators and anticommutators of Clifford algebra elements. This method allows us to find out and prove a number of new properties of Clifford algebra elements.
\end{abstract}
\subjclass{15A66}
\keywords{Clifford algebra, quaternion type, commutator, anticommutator, Lie group, Lie algebra}
\maketitle

\section{Introduction}
In this paper we present a new classification of Clifford algebra elements, which is based on the notion of quaternion type. Using this classification we develop a method for analyzing commutators and anticommutators of Clifford algebra elements. This method allows us to find out and prove a number of new properties of Clifford algebra elements (the main properties are (\ref{1}), (\ref{2})).
In this paper we develop some results of \cite{Marchuk:Shirokov} and \cite{Shirokov}.

\section{Algebras of quaternion type}
Let $\A$ be an $n$-dimensional algebra over the field of complex or real numbers. And let algebra $\A$, considered as an $n$-dimensional vector space, be represented in the form of the direct sum of four vector subspaces
\begin{equation}
\A=\bbE\oplus\bbI\oplus\bbJ\oplus\bbK. \label{A}
\end{equation}
For the elements of these subspaces we use the following designations
$$
\st{\bbE}{A}\in\bbE,\quad\st{\bbI}{B}\in\bbI,\quad\st{\bbE\oplus\bbI}{C}
\in\bbE\oplus\bbI,\ldots
$$
An algebra $\A$ is called {\it the algebra of quaternion type with respect to an operation $\circ:\A\times\A\rightarrow\A$}, if for all elements of considered subspaces the following properties are fulfilled:
\begin{eqnarray}
&&\st{\bbE}{A}\circ\st{\bbE}{B},\ \st{\bbI}{A}\circ\st{\bbI}{B}, \
\st{\bbJ}{A}\circ\st{\bbJ}{B},\
\st{\bbK}{A}\circ\st{\bbK}{B}\in\bbE,\nonumber\\
&&\st{\bbE}{A}\circ\st{\bbI}{B},\ \st{\bbI}{A}\circ\st{\bbE}{B},\
\st{\bbK}{A}\circ\st{\bbJ}{B},\
\st{\bbJ}{A}\circ\st{\bbK}{B}\in\bbI,\label{q:cond}\\
&&\st{\bbE}{A}\circ\st{\bbJ}{B},\ \st{\bbJ}{A}\circ\st{\bbE}{B},\
\st{\bbI}{A}\circ\st{\bbK}{B},\ \st{\bbK}{A}\circ\st{\bbI}{B}\in\bbJ,\nonumber\\
&&\st{\bbE}{A}\circ\st{\bbK}{B},\ \st{\bbK}{A}\circ\st{\bbE}{B},\
\st{\bbI}{A}\circ\st{\bbJ}{B},\ \st{\bbJ}{A}\circ\st{\bbI}{B}\in\bbK.\nonumber
\end{eqnarray}
The operation $\circ$ unessentially should be associative or commutative. This can be seen from the example considered in the following section, where we use the commutator $[\,\cdot\,,\,\cdot\,]$ as the operation $\circ$.

\medskip \noindent

{\bf Elementary examples.} Consider the algebra of quaternions $\mathbb H$. The algebra of quaternions is an associative algebra with the identity element $1$. The elements of $\mathbb H$
(quaternions) can be written in the form
$$
q=a+b i+c j+d k,
$$
where $a,b,c,d\in\mathbb{R}$ and for imaginary units $i,j,k$ the following properties are valid:
\begin{eqnarray*}
&&i^2=j^2=k^2=-1,\\
&&i j=-j i=k,\\
&&j k=-k j=i,\\
&&k i=-i k=j.
\end{eqnarray*}
The algebra $\mathbb H$ is considered as the 4-dimensional vector space that is represented in the form of direct sum of four 1-dimensional vector subspaces $\mathbb{H}=\bbE\oplus\bbI\oplus\bbJ\oplus\bbK$, span over basis elements $1,i,j,k$. Conditions (\ref{q:cond}) are carried out if we use as the operation $\circ$ the usual operation of multiplication of quaternions. Therefore, the algebra of quaternions is an algebra of quaternion type with respect to the operation of multiplication.  \medskip

Another example of algebra of quaternion type is the algebra (set) of smooth complex valued functions of one real variable $x$. For these functions we have multiplication by complex numbers, operations of addition and multiplication (pointwise on $x$). We can represented any function $f$ in the form of the sum of four functions $g_1, h_1, g_2, h_2$ with different properties of parity and a complex accessory:
$$f(x)=f_1(x)+f_2(x)=(g_1(x)+h_1(x))+(g_2(x)+h_2(x)),$$
where
$$f_1(x)=\frac{1}{2}(f(x)+\overline{f}(x)),\quad f_2(x)=\frac{1}{2}(f(x)-\overline{f}(x)),$$
$$g_1(x)=\frac{1}{2}(f_1(x)+f_1(-x))=\frac{1}{4}(f(x)+\overline{f}(x)+f(-x)+\overline{f}(-x)),$$
$$h_1(x)=\frac{1}{2}(f_1(x)-f_1(-x))=\frac{1}{4}(f(x)+\overline{f}(x)-f(-x)-\overline{f}(-x)),$$
$$g_2(x)=\frac{1}{2}(f_2(x)+f_2(-x))=\frac{1}{4}(f(x)-\overline{f}(x)+f(-x)-\overline{f}(-x)),$$
$$h_2(x)=\frac{1}{2}(f_2(x)-f_2(-x))=\frac{1}{4}(f(x)-\overline{f}(x)-f(-x)+\overline{f}(-x)).$$
Let $\bbE, \bbI, \bbJ, \bbK$ be subspaces of all possible functions of the following kinds $g_1(x), h_1(x), g_2(x), h_2(x)$. For the considered functions the conditions (\ref{q:cond}) are carried out, so we can consider the space of functions $f$ as the algebra of quaternion type with respect to the operation of multiplication.
\medskip

In particular, it is possible to consider finite or infinite complex power series of real $x$:
$$\sum_{k=0}^{\infty}(a_k+ib_k)x^k=\sum_{k=0}^{\infty}a_{2k}x^{2k}+
\sum_{k=0}^{\infty}ib_{2k}x^{2k}+\sum_{k=0}^{\infty}a_{2k+1}x^{2k+1}+\sum_{k=0}^{\infty}ib_{2k+1}x^{2k+1}=$$
$$=\st{\bbE}{A}+\st{\bbI}{B}+\st{\bbJ}{C}+\st{\bbK}{D}.$$
Also we can consider analogous sums from $-\infty$ to $\infty$ (we don't consider questions of convergence here).
\bigskip

It follows from (\ref{q:cond}) that any algebra of quaternion type $\A$ has the following subspaces closed with respect to the operation $\circ$:
$$
\bbE,\quad\bbE\oplus\bbI,\quad\bbE\oplus\bbJ,\quad\bbE\oplus\bbK.
$$

Consider elements of algebra $\A$ from different subspaces
\begin{eqnarray}
\bbE,\quad \bbI,\quad \bbJ,\quad \bbK,\quad \bbE\oplus\bbI,\quad \bbE\oplus\bbJ,\quad \bbE\oplus\bbK,\quad \bbI\oplus\bbJ,\quad \bbI\oplus\bbK,\quad \bbJ\oplus\bbK,\label{type}\\
\bbE\oplus\bbI\oplus\bbJ,\quad \bbE\oplus\bbI\oplus\bbK,\quad \bbE\oplus\bbJ\oplus\bbK,\quad \bbI\oplus\bbJ\oplus\bbK,\quad \bbE\oplus\bbI\oplus\bbJ\oplus\bbK=\A.\nonumber
\end{eqnarray}
We say that these elements have different {\it quaternion types} (or {\it types}).

Elements of subspaces $\bbE,\ \bbI,\ \bbJ,\ \bbK$ are called {\it elements of main quaternion types}. Elements of other types are the sums of elements of main quaternion types.

In what follows we omit the sign of direct sum $\oplus$. So $\bbE\bbI\equiv\bbE\oplus\bbI, \quad \bbI\bbJ\bbK\equiv\bbI\oplus\bbJ\oplus\bbK$, etc.

Let's note embeddings of different subspaces (for example, $\bbE\subset\bbE\bbI\subset\bbE\bbI\bbK\subset\bbE\bbI\bbJ\bbK$). The zero element of algebra $\A$ belongs to any quaternion type.

Consider products $A\circ B$ of elements of algebra $\A$ of different quaternion types. We interested in quaternion types of these products.
The following table shows these types:\medskip

\begin{tabular}{|p{0.7cm}|p{0.45cm}p{0.45cm}p{0.45cm}p{0.45cm}p{0.3cm}p{0.3cm}p{0.3cm}p{0.3cm}p{0.3cm}p{0.3cm}p{0.45cm}p{0.45cm}p{0.45cm}p{0.45cm}p{0.2cm}|}
\hline\vsp
 & $\bbE$ & $\bbI$ & $\bbJ$ & $\bbK$ & $\bbE\bbI$ & $\bbE\bbJ$ & $\bbE\bbK$ & $\bbI\bbJ$ & $\bbI\bbK$ & $\bbJ\bbK$ & $\bbE\bbI\bbJ$ & $\bbE\bbI\bbK$ & $\bbE\bbJ\bbK$ & $\bbI\bbJ\bbK$ &$\A$\\ \hline
$\bbE$ & $\bbE$ & $\bbI$ & $\bbJ$ & $\bbK$ & $\bbE\bbI$ & $\bbE\bbJ$ & $\bbE\bbK$ & $\bbI\bbJ$ & $\bbI\bbK$ & $\bbJ\bbK$ & $\bbE\bbI\bbJ$ & $\bbE\bbI\bbK$ & $\bbE\bbJ\bbK$ & $\bbI\bbJ\bbK$ &$\A$\\
$\bbI$ & $\bbI$ & $\bbE$ & $\bbK$ & $\bbJ$ & $\bbE\bbI$ & $\bbI\bbK$ & $\bbI\bbJ$ & $\bbE\bbK$ & $\bbE\bbJ$ & $\bbJ\bbK$ & $\bbE\bbI\bbK$ & $\bbE\bbI\bbJ$ & $\bbI\bbJ\bbK$ & $\bbE\bbJ\bbK$ &$\A$\\
$\bbJ$ & $\bbJ$ & $\bbK$ & $\bbE$ & $\bbI$ & $\bbJ\bbK$ & $\bbE\bbJ$ & $\bbI\bbJ$ & $\bbE\bbK$ & $\bbI\bbK$ & $\bbE\bbI$ & $\bbE\bbJ\bbK$ & $\bbI\bbJ\bbK$ & $\bbE\bbI\bbJ$ & $\bbE\bbI\bbK$ &$\A$\\
$\bbK$ & $\bbK$ & $\bbJ$ & $\bbI$ & $\bbE$ & $\bbJ\bbK$ & $\bbI\bbK$ & $\bbE\bbK$ & $\bbI\bbJ$ & $\bbE\bbJ$ & $\bbE\bbI$ & $\bbI\bbJ\bbK$ & $\bbE\bbJ\bbK$ & $\bbE\bbI\bbK$ & $\bbE\bbI\bbJ$ &$\A$\\
$\bbE\bbI$ & $\bbE\bbI$ & $\bbE\bbI$ & $\bbJ\bbK$ & $\bbJ\bbK$ & $\bbE\bbI$ & $\A$ & $\A$ & $\A$ & $\A$ & $\bbJ\bbK$ & $\A$ & $\A$ & $\A$ & $\A$ &$\A$\\
$\bbE\bbJ$ & $\bbE\bbJ$ & $\bbI\bbK$ & $\bbE\bbJ$ & $\bbI\bbK$ & $\A$ & $\bbE\bbJ$ & $\A$ & $\A$ & $\bbI\bbK$ & $\A$ & $\A$ & $\A$ & $\A$ & $\A$ &$\A$\\
$\bbE\bbK$ & $\bbE\bbK$ & $\bbI\bbJ$ & $\bbI\bbJ$ & $\bbE\bbK$ & $\A$ & $\A$ & $\bbE\bbK$ & $\bbI\bbJ$ & $\A$ & $\A$ & $\A$ & $\A$ & $\A$ & $\A$ &$\A$\\
$\bbI\bbJ$ & $\bbI\bbJ$ & $\bbE\bbK$ & $\bbE\bbK$ & $\bbI\bbJ$ & $\A$ & $\A$ & $\bbI\bbJ$ & $\bbE\bbK$ & $\A$ & $\A$ & $\A$ & $\A$ & $\A$ & $\A$ &$\A$\\
$\bbI\bbK$ & $\bbI\bbK$ & $\bbE\bbJ$ & $\bbI\bbK$ & $\bbE\bbJ$ & $\A$ & $\bbI\bbK$ & $\A$ & $\A$ & $\bbE\bbJ$ & $\A$ & $\A$ & $\A$ & $\A$ & $\A$ &$\A$\\
$\bbJ\bbK$ & $\bbJ\bbK$ & $\bbJ\bbK$ & $\bbE\bbI$ & $\bbE\bbI$ & $\bbJ\bbK$ & $\A$ & $\A$ & $\A$ & $\A$ & $\bbE\bbI$ & $\A$ & $\A$ & $\A$ & $\A$ &$\A$\\
$\bbE\bbI\bbJ$ & $\bbE\bbI\bbJ$ & $\bbE\bbI\bbK$ & $\bbE\bbJ\bbK$ & $\bbI\bbJ\bbK$ & $\A$ & $\A$ & $\A$ & $\A$ & $\A$ & $\A$ & $\A$ & $\A$ & $\A$ & $\A$ &$\A$\\
$\bbE\bbI\bbK$ & $\bbE\bbI\bbK$ & $\bbE\bbI\bbJ$ & $\bbI\bbJ\bbK$ & $\bbE\bbJ\bbK$ & $\A$ & $\A$ & $\A$ & $\A$ & $\A$ & $\A$ & $\A$ & $\A$ & $\A$ & $\A$ &$\A$\\
$\bbE\bbJ\bbK$ & $\bbE\bbJ\bbK$ & $\bbI\bbJ\bbK$ & $\bbE\bbI\bbJ$ & $\bbE\bbI\bbK$ & $\A$ & $\A$ & $\A$ & $\A$ & $\A$ & $\A$ & $\A$ & $\A$ & $\A$ & $\A$ &$\A$\\
$\bbI\bbJ\bbK$ & $\bbI\bbJ\bbK$ & $\bbE\bbJ\bbK$ & $\bbE\bbI\bbK$ & $\bbE\bbI\bbJ$ & $\A$ & $\A$ & $\A$ & $\A$ & $\A$ & $\A$ & $\A$ & $\A$ & $\A$ & $\A$ &$\A$\\
$\A$ & $\A$ & $\A$ & $\A$ & $\A$ & $\A$ & $\A$ & $\A$ & $\A$ & $\A$ & $\A$ & $\A$ & $\A$ & $\A$ & $\A$ & $ \A$ \\ \hline
\end{tabular}

\bigskip

This table illustrates the main properties (\ref{q:cond}) and also, for example, following correspondences:
\begin{eqnarray}
&\st{\bbE\bbI}{U} \ \st{\bbE\bbI}{V},\
\st{\bbJ\bbK}{U} \ \st{\bbJ\bbK}{V}\in\bbE\bbI,\quad
&\st{\bbE\bbI}{U} \ \st{\bbJ\bbK}{V},\
\st{\bbJ\bbK}{U} \ \st{\bbE\bbI}{V}\in\bbJ\bbK,\nonumber\\
&\st{\bbE\bbJ}{U} \ \st{\bbE\bbJ}{V},\
\st{\bbI\bbK}{U} \ \st{\bbI\bbK}{V}\in\bbE\bbJ,\quad
&\st{\bbE\bbJ}{U} \ \st{\bbI\bbK}{V},\
\st{\bbI\bbK}{U} \ \st{\bbE\bbJ}{V}\in\bbI\bbK,\label{main:cond}\\
&\st{\bbE\bbK}{U} \ \st{\bbE\bbK}{V},\
\st{\bbI\bbJ}{U} \ \st{\bbI\bbJ}{V}\in\bbE\bbK,\quad
&\st{\bbE\bbK}{U} \ \st{\bbI\bbJ}{V},\
\st{\bbI\bbJ}{U} \ \st{\bbE\bbK}{V}\in\bbI\bbJ.\nonumber
\end{eqnarray}

Let's note that quaternion typification does not present anything essentially new for the considered examples. But if we look at the Clifford algebra $\cl(p,q)$ from the point of view of quaternion typification, then it is possible to receive substantial and unexpected results as we see in what follows.

%%%%%%%%%%%%%%%%%%%%%%%%%%%%%%%%%%%%%%%%%%%%%%%%%%%%%%%%%%%%%%%%%%%%%%%%%%%%
%%%%%%%%%%%%%%%%%%%%%%%%%%%%%%%%%%%%%%%%%%%%%%%%%%%%%%%%%%%%%%%%%%%%%%%%%%%%

\section{Quaternion typification of Clifford algebra elements}

Let $p, q$ be nonnegative integer numbers and $p+q=n$, $n\geq1$. Consider the real Clifford algebra $\cl^\R(p,q)$ or the complex Clifford algebra $\cl^\C(p,q)$. In the case when results are true for both cases, we write $\cl(p,q)$. The construction of Clifford algebra $\cl(p,q)$ is discussed in details in \cite{Lounesto} or \cite{Marchuk:Shirokov}. Let $e$ be the identity element and let $e^a$, $a=1,\ldots,n$ be generators of the Clifford algebra $\cl(p,q)$,
$$
e^a e^b+e^b e^a=2\eta^{ab}e,
$$
where $\eta=||\eta^{ab}||$ is the diagonal matrix with $p$ pieces of $+1$ and $q$ pieces of $-1$ on the diagonal. Elements
$$
e^{a_1\ldots a_k}=e^{a_1}\ldots e^{a_k},\quad a_1<\ldots<a_k,\,k=1,\ldots,n,
$$
together with the identity element $e$, form the basis of the Clifford
algebra. The number of basis elements is equal to $2^n$.

We denote by $\cl_k(p,q)$ the vector spaces that span over the basis elements
$e^{a_1\ldots a_k}$. Elements of $\cl_k(p,q)$ are said to be
elements of rank $k$. Sometimes we denote elements of rank $k$ by $\st{k}{W}, \st{k}{V}, \ldots$ We have the following classification of Clifford algebra elements based on the notion of rank:
\begin{eqnarray}
\cl(p,q)=\oplus_{k=0}^{n}\cl_k(p,q).\label{ranks}
\end{eqnarray}
So, any Clifford algebra element is an element of some rank or a sum of elements of different ranks:
\begin{eqnarray}
U=\st{k_1}{U}+\st{k_2}{U}+\ldots+\st{k_m}{U},\qquad 0\leq k_1<\ldots<k_m\leq n.
\end{eqnarray}

Also we have classification of Clifford algebra elements based on the notion of parity (Clifford algebra as a superalgebra):
\begin{eqnarray}
\cl(p,q)=\cl_{even}(p,q)\oplus\cl_{odd}(p,q),\label{evenness}
\end{eqnarray}
where
$$\cl_{even}(p,q)=\cl_0(p,q)\oplus\cl_2(p,q)\oplus\cl_4(p,q)\oplus\ldots$$
$$\cl_{odd}(p,q)=\cl_1(p,q)\oplus\cl_3(p,q)\oplus\cl_5(p,q)\oplus\ldots$$

Denote by $[U,V]$ the commutator and  by $\{U,V\}$ the anticommutator of
Clifford algebra elements
$$
[U,V]=UV-VU,\quad \{U,V\}=UV+VU
$$
and note that
\begin{eqnarray}
UV=\frac{1}{2}[U,V]+\frac{1}{2}\{U,V\}.\label{proizv}
\end{eqnarray}

\bigskip

Let us consider the Clifford algebra as the vector space and represent it in the form of the direct sum of four subspaces:
\begin{equation}
\cl(p,q)=\cl_{\overline 0}(p,q)\oplus\cl_{\overline 1}(p,q)\oplus
\cl_{\overline 2}(p,q)\oplus\cl_{\overline 3}(p,q),\label{kv}
\end{equation}
where
\begin{eqnarray*}
\cl_{\overline
0}(p,q)&=&\cl_0(p,q)\oplus\cl_4(p,q)\oplus\cl_8(p,q)\oplus\ldots,\\
\cl_{\overline
1}(p,q)&=&\cl_1(p,q)\oplus\cl_5(p,q)\oplus\cl_9(p,q)\oplus\ldots,\\
\cl_{\overline
2}(p,q)&=&\cl_2(p,q)\oplus\cl_6(p,q)\oplus\cl_{10}(p,q)\oplus\ldots,\\
\cl_{\overline
3}(p,q)&=&\cl_3(p,q)\oplus\cl_7(p,q)\oplus\cl_{11}(p,q)\oplus\ldots
\end{eqnarray*}
and in the right hand parts there are direct sums of subspaces with dimensions differ on 4. We suppose that $\cl_k(p,q)=\emptyset$ for $k>p+q$.

If $\st{\overline k}{U}\in\cl_{\overline k}(p,q)$, then we may write
$$
\st{\overline k}{U}=\st{k}{U}+\st{k+4}{U}+\st{k+8}{U}+\ldots, \qquad
k=0,1,2,3.
$$

Note that the Clifford algebra $\cl(p,q)$ with Clifford product doesn't form an algebra of quaternion type. However, we have the following theorem.
\begin{thm}
a) The Clifford algebra $\cl(p,q)$ is an algebra of quaternion type with respect to the operation $\quad U, V \rightarrow \{U,V\}.$
In designations of the previous section we have
$$\quad \bbE=\cl_{\overline 0}(p,q),\quad\bbI=\cl_{\overline1}(p,q),\quad\bbJ=\cl_{\overline 2}(p,q),\quad\bbK=\cl_{\overline 3}(p,q)\quad.$$
b) The Clifford algebra $\cl(p,q)$ is an algebra of quaternion type with respect to the operation $\quad U, V \rightarrow [U,V].$
We have
$$\quad \bbE=\cl_{\overline 2}(p,q),\quad\bbI=\cl_{\overline 3}(p,q),\quad\bbJ=\cl_{\overline 0}(p,q),\quad\bbK=\cl_{\overline 1}(p,q)\quad.$$
\end{thm}

{\bf Remark.} Statements of the theorem are equivalent to the following properties:

for any two Clifford algebra elements $U, V$ from given subspaces, there exists Clifford algebra element $W$ such that
\begin{eqnarray}
&&[\st{\overline k}{U},\st{\overline k}{V}]=\st{\overline 2}{W},\qquad k=0, 1, 2, 3 \nonumber;\\
&&[\st{\overline k}{U},\st{\overline 2}{V}]=\st{\overline k}{W}, \qquad k=0, 1, 2, 3 \nonumber; \\
&&[\st{\overline 0}{U},\st{\overline 1}{V}]=\st{\overline 3}{W}, \quad  [\st{\overline 0}{U},\st{\overline 3}{V}]=\st{\overline 1}{W}, \quad [\st{\overline 1}{U},\st{\overline 3}{V}]=\st{\overline 0}{W} \nonumber,
\end{eqnarray}

\begin{eqnarray}
&&\{\st{\overline k}{U},\st{\overline k}{V}\}=\st{\overline 0}{W},\qquad k=0, 1, 2, 3 \nonumber;\\
&&\{\st{\overline k}{U},\st{\overline 0}{V}\}=\st{\overline k}{W}, \qquad k=0, 1, 2, 3 \nonumber; \\
&&\{\st{\overline 1}{U},\st{\overline 2}{V}\}=\st{\overline 3}{W},  \quad \{\st{\overline 1}{U},\st{\overline 3}{V}\}=\st{\overline 2}{W}, \quad \{\st{\overline 2}{U},\st{\overline 3}{V}\}=\st{\overline 1}{W}\nonumber.
\end{eqnarray}

These formulas allowed us to define subspaces of Clifford algebra commutators and anticommutators.

Let's write down these properties in the other notation:
\begin{eqnarray}
&&[\overline k,\overline k]\subseteq\overline{\textbf{2}},\qquad k=0, 1, 2, 3 \nonumber;\\
&&[\overline k,\overline 2]\subseteq\overline{\textbf{k}}, \qquad k=0, 1, 2, 3 \label{1}; \\
&&[\overline 0,\overline 1]\subseteq\overline{\textbf{3}}, \quad  [\overline 0,\overline 3]\subseteq\overline{\textbf{1}}, \quad [\overline 1,\overline 3]\subseteq\overline{\textbf{0}} \nonumber,
\end{eqnarray}

\begin{eqnarray}
&&\{\overline k,\overline k\}\subseteq\overline{\textbf{0}},\qquad k=0, 1, 2, 3 \nonumber;\\
&&\{\overline k,\overline 0\}\subseteq\overline{\textbf{k}}, \qquad k=0, 1, 2, 3; \label{2} \\
&&\{\overline 1,\overline 2\}\subseteq\overline{\textbf{3}},  \quad \{\overline 1,\overline 3\}\subseteq\overline{\textbf{2}}, \quad \{\overline 2,\overline 3\}\subseteq\overline{\textbf{1}}\nonumber,
\end{eqnarray}
where we denote $\cl_{\overline k}(p,q)$ by $\overline{\textbf{k}}$ and any element $\st{\overline k}{U}\in\cl_{\overline k}(p,q)$ by $\overline{k}$. Note that the first and the second $\overline k$ in $[\overline k,\overline k]$ (or $\{\overline k,\overline k\}$) are two different Clifford algebra elements from the same subspace $\overline{\textbf{k}}$.

\proof. \, To prove these statements we use results of the paper \cite{Shirokov} (theorems 1 and 2). Namely, for $n\geq k\geq l$ we have
$$
[\st{k}{U},\st{l}{V}]=\left\lbrace
\begin{array}{ll}
\st{k-l}{W}+\st{k-l+4}{W}+\ldots, & \mbox{\rm if $k$ -
even, $l$ - odd;}\\
\st{k-l+2}{W}+\st{k-l+6}{W}+\ldots, & \mbox{\rm other cases,}
\end{array}
\right.
$$
$$
\{\st{k}{U},\st{l}{V}\}=\left\lbrace
\begin{array}{ll}
\st{k-l+2}{W}+\st{k-l+6}{W}+\ldots, & \mbox{\rm if $k$ -
even, $l$ - odd;}\\
\st{k-l}{W}+\st{k-l+4}{W}+\ldots, & \mbox{\rm other cases,}
\end{array}
\right.
$$
where the series are finite (but now it does not matter for us).

Let's write down some special cases of these formulas for natural $t\geq s\geq 0$.
$$[\st{4t}{U},\st{4s}{V}]=[\st{1+4t}{U},\st{1+4s}{V}]=[\st{2+4t}{U},\st{2+4s}{V}]=[\st{3+4t}{U},\st{3+4s}{V}]=\st{4(t-s)+2}{W}+\st{4(t-s)+6}{W}+\ldots,$$
$$[\st{1+4t}{U},\st{4s}{V}]=[\st{3+4t}{U},\st{2+4s}{V}]=\st{4(t-s)+3}{W}+\st{4(t-s)+7}{W}+\ldots,$$
$$[\st{2+4t}{U},\st{4s}{V}]=[\st{3+4t}{U},\st{1+4s}{V}]=\st{4(t-s)+4}{W}+\st{4(t-s)+8}{W}+\ldots,$$
$$[\st{3+4t}{U},\st{4s}{V}]=\st{4(t-s)+5}{W}+\st{4(t-s)+9}{W}+\ldots,$$
$$[\st{2+4t}{U},\st{1+4s}{V}]=\st{4(t-s)+1}{W}+\st{4(t-s)+5}{W}+\ldots,$$

$$\{\st{4t}{U},\st{4s}{V}\}=\{\st{1+4t}{U},\st{1+4s}{V}\}=\{\st{2+4t}{U},\st{2+4s}{V}\}=\{\st{3+4t}{U},\st{3+4s}{V}\}=\st{4(t-s)}{W}+\st{4(t-s)+4}{W}+\ldots,$$
$$\{\st{1+4t}{U},\st{4s}{V}\}=\{\st{3+4t}{U},\st{2+4s}{V}\}=\st{4(t-s)+1}{W}+\st{4(t-s)+5}{W}+\ldots,$$
$$\{\st{2+4t}{U},\st{4s}{V}\}=\{\st{3+4t}{U},\st{1+4s}{V}\}=\st{4(t-s)+2}{W}+\st{4(t-s)+6}{W}+\ldots,$$
$$\{\st{3+4t}{U},\st{4s}{V}\}=\{\st{2+4t}{U},\st{1+4s}{V}\}=\st{4(t-s)+3}{W}+\st{4(t-s)+7}{W}+\ldots$$
And for natural $s > t\geq 0$ we have
$$[\st{4t}{U},\st{4s}{V}]=[\st{1+4t}{U},\st{1+4s}{V}]=[\st{2+4t}{U},\st{2+4s}{V}]=[\st{3+4t}{U},\st{3+4s}{V}]=\st{4(s-t)+2}{W}+\st{4(s-t)+6}{W}+\ldots,$$
$$[\st{1+4t}{U},\st{4s}{V}]=[\st{3+4t}{U},\st{2+4s}{V}]=\st{4(s-t)-1}{W}+\st{4(s-t)+3}{W}+\ldots,$$
$$[\st{2+4t}{U},\st{4s}{V}]=[\st{3+4t}{U},\st{1+4s}{V}]=\st{4(s-t)}{W}+\st{4(s-t)+4}{W}+\ldots,$$
$$[\st{3+4t}{U},\st{4s}{V}]=\st{4(s-t)-3}{W}+\st{4(s-t)+1}{W}+\ldots,$$
$$[\st{2+4t}{U},\st{1+4s}{V}]=\st{4(s-t)+1}{W}+\st{4(s-t)+5}{W}+\ldots,$$

$$\{\st{4t}{U},\st{4s}{V}\}=\{\st{1+4t}{U},\st{1+4s}{V}\}=\{\st{2+4t}{U},\st{2+4s}{V}\}=\{\st{3+4t}{U},\st{3+4s}{V}\}=\st{4(s-t)}{W}+\st{4(s-t)+4}{W}+\ldots,$$
$$\{\st{1+4t}{U},\st{4s}{V}\}=\{\st{3+4t}{U},\st{2+4s}{V}\}=\st{4(s-t)+1}{W}+\st{4(s-t)+5}{W}+\ldots,$$
$$\{\st{2+4t}{U},\st{4s}{V}\}=\{\st{3+4t}{U},\st{1+4s}{V}\}=\st{4(s-t)-2}{W}+\st{4(s-t)+2}{W}+\ldots,$$
$$\{\st{3+4t}{U},\st{4s}{V}\}=\{\st{2+4t}{U},\st{1+4s}{V}\}=\st{4(s-t)-1}{W}+\st{4(s-t)+3}{W}+\ldots$$
From these formulas we get the statements of the theorem.
$\blacksquare$

\bigskip

We use the following notations:
$$
\cl_{\overline{kl}}(p,q)=\cl_{\overline
k}(p,q)\oplus\cl_{\overline l}(p,q),\quad 0\leq k<l\leq 3.
$$
$$
\cl_{\overline{klm}}(p,q)=\cl_{\overline
k}(p,q)\oplus\cl_{\overline l}(p,q)\oplus\cl_{\overline
m}(p,q),\quad 0\leq k<l<m\leq 3.
$$
If $\st{\overline{kl}}{U}\in\cl_{\overline{kl}}(p,q)$, then
$$
\st{\overline{kl}}{U}=\st{\overline{k}}{U}+\st{\overline{l}}{U}=
(\st{k}{U}+\st{l}{U})+(\st{k+4}{U}+\st{l+4}{U})+\ldots,\quad
0\leq k<l\leq 3.
$$

Consider elements of the Clifford algebra $\cl(p,q)$ from different subspaces
\begin{eqnarray}
\cl_{\overline 0}(p,q),\quad \cl_{\overline 1}(p,q),\quad \cl_{\overline 2}(p,q),\quad \cl_{\overline 3}(p,q),\quad \cl_{\overline {01}}(p,q),\quad \cl_{\overline {02}}(p,q),\nonumber \\
\cl_{\overline {03}}(p,q),\quad \cl_{\overline {12}}(p,q), \quad \cl_{\overline {13}}(p,q),\quad \cl_{\overline {23}}(p,q),\quad \cl_{\overline {012}}(p,q),\label{tip}\\
\cl_{\overline {013}}(p,q),\quad \cl_{\overline {023}}(p,q),\quad \cl_{\overline {123}}(p,q),\quad
\cl_{\overline {0123}}(p,q)=\cl(p,q).\nonumber
\end{eqnarray}
Then we say that these elements have different {\it quaternion types} (or {\it types}).

Elements of subspaces $\cl_{\overline 0}(p,q),\, \cl_{\overline 1}(p,q),\, \cl_{\overline 2}(p,q),\, \cl_{\overline 3}(p,q)$ are called {\it elements of main quaternion types}. Elements of other types are represented in the form of sums of elements of main quaternion types. Suppose that the zero element of the Clifford algebra $\cl(p,q)$ belongs to any quaternion type.

\bigskip
The classification of all Clifford algebra $\cl(p,q)$ elements (for any integer nonnegative numbers $p+q=n$) into 15 quaternion types (\ref{tip}) and the application of Theorem 1 to calculating the quaternion types of commutators and anticommutators of Clifford algebra elements constitute the essence of the method of quaternion typification of Clifford algebra elements.
\bigskip

There are two well-known classifications of Clifford algebra elements: classification based on the notion of rank (\ref{ranks}) and classification based on the notion of parity (\ref{evenness}). Often we consider expressions of Clifford algebra elements and we are interested in rank of these expressions. We can consider the same question based on the notion of quaternion type and receive substantial results. In some questions the classification of Clifford algebra elements based on the notion of quaternion type is more suitable than the classification based on the notion of rank for Clifford algebras of dimensions $n\geq 4$. Formulas (\ref{1}) and (\ref{2}) allowed us to define quaternion type of commutators and anticommutators of Clifford algebra elements of given quaternion types.

For example, for Clifford algebra of dimension $n=4$ we have the following two equivalent expressions:
$$[\st{3}{U},\st{3}{V}] \in\cl_2(p,q),\quad n=4;$$
$$[\st{\overline{3}}{U},\st{\overline{3}}{V}]\in\cl_{\overline 2}(p,q),\quad n=4.$$
But for $n=20$ the following two expressions represent the same property:
\begin{eqnarray}
&&[\st{2}{U},\st{2}{V}],\quad[\st{6}{U},\st{6}{V}],\quad[\st{10}{U},\st{10}{V}],\quad[\st{2}{U},\st{6}{V}],\quad[\st{6}{U},\st{2}{V}],\quad[\st{2}{U},\st{10}{V}],\quad[\st{10}{U},\st{2}{V}],\quad[\st{6}{U},\st{10}{V}],\nonumber\\ &&[\st{10}{U},\st{6}{V}]\in\cl_2(p,q)\oplus\cl_6(p,q)\oplus\cl_{10}(p,q)\oplus\cl_{14}(p,q)\oplus\cl_{18}(p,q),\quad n=20;\nonumber
\end{eqnarray}

$$[\st{\overline{2}}{U},\st{\overline{2}}{V}]\in\cl_{\overline 2}(p,q),\quad n=20.$$

So, for $n\geq 4$ the classification based on the notion of type is more suitable, but it is rougher than the classification based on the notion of rank. The classification based on the notion of parity is the most roughest among these three classifications.

Let's represent results of Theorem 1 in the form of tables, which display action of commutators and anticommutators on elements of the Clifford algebra of different quaternion types. By $\A$ denote the Clifford algebra $\cl(p,q)=\cl_{\overline 0\overline 1\overline 2\overline 3}(p,q)$.

These tables illustrate the main properties (\ref{1}), (\ref{2}) and also, for example, following correspondences:
\begin{eqnarray}
&&[\overline{01},\overline{01}],\, [\overline{23},\overline{23}]\subseteq\overline{\textbf{23}},\qquad [\overline{01},\overline{23}],\, [\overline{23},\overline{01}]\subseteq\overline{\textbf{01}}\nonumber\\
&&[\overline{02},\overline{02}],\, [\overline{13},\overline{13}]\subseteq\overline{\textbf{02}},\qquad [\overline{02},\overline{13}],\, [\overline{13},\overline{02}]\subseteq\overline{\textbf{13}}\\
&&[\overline{03},\overline{03}],\, [\overline{12},\overline{12}]\subseteq\overline{\textbf{12}},\qquad [\overline{03},\overline{12}],\, [\overline{12},\overline{03}]\subseteq\overline{\textbf{03}},\nonumber
\end{eqnarray}
\begin{eqnarray}
&&\{\overline{01},\overline{01}\},\, \{\overline{23},\overline{23}\}\subseteq\overline{\textbf{01}},\qquad \{\overline{01},\overline{23}\},\, \{\overline{23},\overline{01}\}\subseteq\overline{\textbf{23}}\nonumber\\
&&\{\overline{02},\overline{02}\},\, \{\overline{13},\overline{13}\}\subseteq\overline{\textbf{02}},\qquad \{\overline{02},\overline{13}\},\, \{\overline{13},\overline{02}\}\subseteq\overline{\textbf{13}}\\
&&\{\overline{03},\overline{03}\},\, \{\overline{12},\overline{12}\}\subseteq\overline{\textbf{03}},\qquad \{\overline{03},\overline{12}\},\, \{\overline{12},\overline{03}\}\subseteq\overline{\textbf{12}}.\nonumber
\end{eqnarray}

\medskip

\begin{tabular}{|p{0.5cm}|p{0.4cm} p{0.4cm} p{0.4cm} p{0.4cm} p{0.3cm} p{0.3cm} p{0.3cm} p{0.3cm} p{0.3cm} p{0.3cm} p{0.4cm} p{0.4cm} p{0.4cm} p{0.4cm} p{0.4cm}|}\hline\vsp
$[,]$ & $\overline 0$ & $\overline 1$ & $\overline 2$ & $\overline 3$ & $\overline 0\overline 1$ & $\overline 0\overline 2$ & $\overline 0\overline 3$ & $\overline 1\overline 2$ & $\overline 1\overline 3$ & $\overline 2\overline 3$ & $\overline 0\overline 1\overline 2$ & $\overline 0\overline 1\overline 3$ & $\overline 0\overline 2\overline 3$ & $\overline 1\overline 2\overline 3$ &$\A$\\ \hline\vsp
$\overline 0$ & $\overline 2$ & $\overline 3$ & $\overline 0$ & $\overline 1$ & $\overline 2\overline 3$ & $\overline 0\overline 2$ & $\overline 1\overline 2$ & $\overline 0\overline 3$ & $\overline 1\overline 3$ & $\overline 0\overline 1$ & $\overline 0\overline 2\overline 3$ & $\overline 1\overline 2\overline 3$ & $\overline 0\overline 1\overline 2$ & $\overline 0\overline 1\overline 3$ &$\A$\\
$\overline 1$ & $\overline3$ & $\overline 2$ & $\overline 1$ & $\overline 0$ & $\overline 2\overline 3$ & $\overline 1\overline 3$ & $\overline 0\overline 3$ & $\overline 1\overline 2$ & $\overline 0\overline 2$ & $\overline 0\overline 1$ & $\overline 1\overline 2\overline 3$ & $\overline 0\overline 2\overline 3$ & $\overline 0\overline 1\overline 3$ & $\overline 0\overline 1\overline 2$ &$\A$\\
$\overline 2$ & $\overline 0$ & $\overline 1$ & $\overline 2$ & $\overline 3$ & $\overline 0\overline 1$ & $\overline 0\overline 2$ & $\overline 0\overline 3$ & $\overline 1\overline 2$ & $\overline 1\overline 3$ & $\overline 2\overline 3$ & $\overline 0\overline 1\overline 2$ & $\overline 0\overline 1\overline 3$ & $\overline 0\overline 2\overline 3$ & $\overline 1\overline 2\overline 3$ &$\A$\\
$\overline 3$ & $\overline 1$ & $\overline 0$ & $\overline 3$ & $\overline 2$ & $\overline 0\overline 1$ & $\overline 1\overline 3$ & $\overline 1\overline 2$ & $\overline 0\overline 3$ & $\overline 0\overline 2$ & $\overline 2\overline 3$ & $\overline 0\overline 1\overline 3$ & $\overline 0\overline 1\overline 2$ & $\overline 1\overline 2\overline 3$ & $\overline 0\overline 2\overline 3$ &$\A$\\
$\overline 0\overline 1$ &$\overline 2\overline 3$ & $\overline 2\overline 3$ & $\overline 0\overline 1$ & $\overline 0\overline 1$ & $\overline 2\overline 3$ & $\A$ & $\A$ & $\A$ & $\A$ & $\overline 0\overline 1$ & $\A$ & $\A$ & $\A$ & $\A$ &$\A$\\
$\overline 0\overline 2$ & $\overline 0\overline2$ & $\overline 1\overline 3$ & $\overline 0\overline 2$ & $\overline 1\overline 3$ & $\A$ & $\overline 0\overline 2$ & $\A$ & $\A$ & $\overline 1\overline 3$ & $\A$ & $\A$ & $\A$ & $\A$ & $\A$ &$\A$\\
$\overline 0\overline 3$ & $\overline 1\overline2$ & $\overline 0\overline 3$ & $\overline 0\overline 3$ & $\overline 1\overline 2$ & $\A$ & $\A$ & $\overline 1\overline 2$ & $\overline 0\overline 3$ & $\A$ & $\A$ & $\A$ & $\A$ & $\A$ & $\A$ &$\A$\\
$\overline 1\overline 2$ & $\overline 0\overline3$ & $\overline 1\overline 2$ & $\overline 1\overline 2$ & $\overline 0\overline 3$ & $\A$ & $\A$ & $\overline 0\overline 3$ & $\overline 1\overline 2$ & $\A$ & $\A$ & $\A$ & $\A$ & $\A$ & $\A$ &$\A$\\
$\overline 1\overline 3$ & $\overline 1\overline3$ & $\overline 0\overline 2$ & $\overline 1\overline 3$ & $\overline 0\overline 2$ & $\A$ & $\overline 1\overline 3$ & $\A$ & $\A$ & $\overline 0\overline 2$ & $\A$ & $\A$ & $\A$ & $\A$ & $\A$ &$\A$\\
$\overline 2\overline 3$ & $\overline 0\overline 1$ & $\overline 0\overline 1$ & $\overline 2\overline 3$ & $\overline 2\overline 3$ & $\overline 0\overline 1$ & $\A$ & $\A$ & $\A$ & $\A$ & $\overline 2\overline 3$ & $\A$ & $\A$ & $\A$ & $\A$ &$\A$\\
$\overline 0\overline 1\overline 2$ & $\overline0\overline 2\overline 3$ & $\overline 1\overline 2\overline 3$ & $\overline 0\overline 1\overline 2$ & $\overline 0\overline 1\overline 3$ & $\A$ & $\A$ & $\A$ & $\A$ & $\A$ & $\A$ & $\A$ & $\A$ & $\A$ & $\A$ &$\A$\\
$\overline 0\overline 1\overline 3$ & $\overline1\overline 2\overline 3$ & $\overline 0\overline 2\overline 3$ & $\overline 0\overline 1\overline 3$ & $\overline 0\overline 1\overline 2$ & $\A$ & $\A$ & $\A$ & $\A$ & $\A$ & $\A$ & $\A$ & $\A$ & $\A$ & $\A$ &$\A$\\
$\overline 0\overline 2\overline 3$ & $\overline 0\overline 1\overline 2$ & $\overline 0\overline 1\overline 3$ & $\overline 0\overline 2\overline 3$ & $\overline 1\overline 2\overline 3$ & $\A$ & $\A$ & $\A$ & $\A$ & $\A$ & $\A$ & $\A$ & $\A$ & $\A$ & $\A$ &$\A$\\
$\overline 1\overline 2\overline 3$ & $\overline 0\overline 1\overline 3$ & $\overline 0\overline 1\overline 2$ & $\overline 1\overline 2\overline 3$ & $\overline 0\overline 2\overline 3$ & $\A$ & $\A$ & $\A$ & $\A$ & $\A$ & $\A$ & $\A$ & $\A$ & $\A$ & $\A$ &$\A$\\
$\A$ & $\A$ & $\A$ & $\A$ & $\A$ & $\A$ & $\A$ & $\A$ & $\A$ & $\A$ & $\A$ & $\A$ & $\A$ & $\A$ & $\A$ & $ \A$ \\ \hline
\end{tabular}
\bigskip

\begin{tabular}{|p{0.5cm}|p{0.4cm} p{0.4cm} p{0.4cm} p{0.4cm} p{0.3cm} p{0.3cm} p{0.3cm} p{0.3cm} p{0.3cm} p{0.3cm} p{0.4cm} p{0.4cm} p{0.4cm} p{0.4cm} p{0.4cm}|}\hline\vsp
$\{,\}$ & $\overline 0$ & $\overline 1$ & $\overline 2$ & $\overline 3$ & $\overline 0\overline 1$ & $\overline 0\overline 2$ & $\overline 0\overline 3$ & $\overline 1\overline 2$ & $\overline 1\overline 3$ & $\overline 2\overline 3$ & $\overline 0\overline 1\overline 2$ & $\overline 0\overline 1\overline 3$ & $\overline 0\overline 2\overline 3$ & $\overline 1\overline 2\overline 3$ &$\A$\\ \hline\vsp
$\overline 0$ & $\overline 0$ & $\overline 1$ & $\overline 2$ & $\overline 3$ & $\overline 0\overline 1$ & $\overline 0\overline 2$ & $\overline 0\overline 3$ & $\overline 1\overline 2$ & $\overline 1\overline 3$ & $\overline 2\overline 3$ & $\overline 0\overline 1\overline 2$ & $\overline 0\overline 1\overline 3$ & $\overline 0\overline 2\overline 3$ & $\overline 1\overline 2\overline 3$ &$\A$\\
$\overline 1$ & $\overline 1$ & $\overline 0$ & $\overline 3$ & $\overline 2$ & $\overline 0\overline 1$ & $\overline 1\overline 3$ & $\overline 1\overline 2$ & $\overline 0\overline 3$ & $\overline 0\overline 2$ & $\overline 2\overline 3$ & $\overline 0\overline 1\overline 3$ & $\overline 0\overline 1\overline 2$ & $\overline 1\overline 2\overline 3$ & $\overline 0\overline 2\overline 3$ &$\A$\\
$\overline 2$ & $\overline 2$ & $\overline 3$ & $\overline 0$ & $\overline 1$ & $\overline 2\overline 3$ & $\overline 0\overline 2$ & $\overline 1\overline 2$ & $\overline 0\overline 3$ & $\overline 1\overline 3$ & $\overline 0\overline 1$ & $\overline 0\overline 2\overline 3$ & $\overline 1\overline 2\overline 3$ & $\overline 0\overline 1\overline 2$ & $\overline 0\overline 1\overline 3$ &$\A$\\
$\overline 3$ & $\overline 3$ & $\overline 2$ & $\overline 1$ & $\overline 0$ & $\overline 2\overline 3$ & $\overline 1\overline 3$ & $\overline 0\overline 3$ & $\overline 1\overline 2$ & $\overline 0\overline 2$ & $\overline 0\overline 1$ & $\overline 1\overline 2\overline 3$ & $\overline 0\overline 2\overline 3$ & $\overline 0\overline 1\overline 3$ & $\overline 0\overline 1\overline 2$ &$\A$\\
$\overline 0\overline 1$ & $\overline 0\overline 1$ & $\overline 0\overline 1$ & $\overline 2\overline 3$ & $\overline 2\overline 3$ & $\overline 0\overline 1$ & $\A$ & $\A$ & $\A$ & $\A$ & $\overline 2\overline 3$ & $\A$ & $\A$ & $\A$ & $\A$ &$\A$\\
$\overline 0\overline 2$ & $\overline 0\overline 2$ & $\overline 1\overline 3$ & $\overline 0\overline 2$ & $\overline 1\overline 3$ & $\A$ & $\overline 0\overline 2$ & $\A$ & $\A$ & $\overline 1\overline 3$ & $\A$ & $\A$ & $\A$ & $\A$ & $\A$ &$\A$\\
$\overline 0\overline 3$ & $\overline 0\overline 3$ & $\overline 1\overline 2$ & $\overline 1\overline 2$ & $\overline 0\overline 3$ & $\A$ & $\A$ & $\overline 0\overline 3$ & $\overline 1\overline 2$ & $\A$ & $\A$ & $\A$ & $\A$ & $\A$ & $\A$ &$\A$\\
$\overline 1\overline 2$ & $\overline 1\overline 2$ & $\overline 0\overline 3$ & $\overline 0\overline 3$ & $\overline 1\overline 2$ & $\A$ & $\A$ & $\overline 1\overline 2$ & $\overline 0\overline 3$ & $\A$ & $\A$ & $\A$ & $\A$ & $\A$ & $\A$ &$\A$\\
$\overline 1\overline 3$ & $\overline 1\overline 3$ & $\overline 0\overline 2$ & $\overline 1\overline 3$ & $\overline 0\overline 2$ & $\A$ & $\overline 1\overline 3$ & $\A$ & $\A$ & $\overline 0\overline 2$ & $\A$ & $\A$ & $\A$ & $\A$ & $\A$ &$\A$\\
$\overline 2\overline 3$ & $\overline 2\overline 3$ & $\overline 2\overline 3$ & $\overline 0\overline 1$ & $\overline 0\overline 1$ & $\overline 2\overline 3$ & $\A$ & $\A$ & $\A$ & $\A$ & $\overline 0\overline 1$ & $\A$ & $\A$ & $\A$ & $\A$ &$\A$\\
$\overline 0\overline 1\overline 2$ & $\overline 0\overline 1\overline 2$ & $\overline 0\overline 1\overline 3$ & $\overline 0\overline 2\overline 3$ & $\overline 1\overline 2\overline 3$ & $\A$ & $\A$ & $\A$ & $\A$ & $\A$ & $\A$ & $\A$ & $\A$ & $\A$ & $\A$ &$\A$\\
$\overline 0\overline 1\overline 3$ & $\overline 0\overline 1\overline 3$ & $\overline 0\overline 1\overline 2$ & $\overline 1\overline 2\overline 3$ & $\overline 0\overline 2\overline 3$ & $\A$ & $\A$ & $\A$ & $\A$ & $\A$ & $\A$ & $\A$ & $\A$ & $\A$ & $\A$ &$\A$\\
$\overline 0\overline 2\overline 3$ & $\overline 0\overline 2\overline 3$ & $\overline 1\overline 2\overline 3$ & $\overline 0\overline 1\overline 2$ & $\overline 0\overline 1\overline 3$ & $\A$ & $\A$ & $\A$ & $\A$ & $\A$ & $\A$ & $\A$ & $\A$ & $\A$ & $\A$ &$\A$\\
$\overline 1\overline 2\overline 3$ & $\overline 1\overline 2\overline 3$ & $\overline 0\overline 2\overline 3$ & $\overline 0\overline 1\overline 3$ & $\overline 0\overline 1\overline 2$ & $\A$ & $\A$ & $\A$ & $\A$ & $\A$ & $\A$ & $\A$ & $\A$ & $\A$ & $\A$ &$\A$\\
$\A$ & $\A$ & $\A$ & $\A$ & $\A$ & $\A$ & $\A$ & $\A$ & $\A$ & $\A$ & $\A$ & $\A$ & $\A$ & $\A$ & $\A$ & $ \A$ \\ \hline
\end{tabular}

\bigskip

Using (\ref{proizv}) we get the table for products of two Clifford algebra elements of different types:

\bigskip
\begin{tabular}{|p{0.5cm}|p{0.4cm} p{0.4cm} p{0.4cm} p{0.4cm} p{0.3cm} p{0.3cm} p{0.3cm} p{0.3cm} p{0.3cm} p{0.3cm} p{0.4cm} p{0.4cm} p{0.4cm} p{0.4cm} p{0.4cm}|}\hline\vsp
& $\overline 0$ & $\overline 1$ & $\overline 2$ & $\overline 3$ & $\overline 0\overline 1$ & $\overline 0\overline 2$ & $\overline 0\overline 3$ & $\overline 1\overline 2$ & $\overline 1\overline 3$ & $\overline 2\overline 3$ & $\overline 0\overline 1\overline 2$ & $\overline 0\overline 1\overline 3$ & $\overline 0\overline 2\overline 3$ & $\overline 1\overline 2\overline 3$ &$\A$\\ \hline\vsp
$\overline 0$ & $\overline {02}$ & $\overline {13}$ & $\overline {02}$ & $\overline {13}$ & $\A$ & $\overline {02}$ & $\A$ & $\A$ & $\overline {13}$ & $\A$ & $\A$ & $\A$ & $\A$ & $\A$ &$\A$\\
$\overline 1$ & $\overline {13}$ & $\overline {02}$ & $\overline {13}$ & $\overline {02}$ & $\A$ & $\overline {13}$ & $\A$ & $\A$ & $\overline {02}$ & $\A$ & $\A$ & $\A$ & $\A$ & $\A$ &$\A$\\
$\overline 2$ & $\overline {02}$ & $\overline {13}$ & $\overline {02}$ & $\overline {13}$ & $\A$ & $\overline {02}$ & $\A$ & $\A$ & $\overline {13}$ & $\A$ & $\A$ & $\A$ & $\A$ & $\A$ &$\A$\\
$\overline 3$ & $\overline {13}$ & $\overline {02}$ & $\overline {13}$ & $\overline {02}$ & $\A$ & $\overline {13}$ & $\A$ & $\A$ & $\overline {02}$ & $\A$ & $\A$ & $\A$ & $\A$ & $\A$ &$\A$\\
$\overline {01}$ & $\A$ & $A$ & $\A$ & $\A$ & $\A$ & $\A$ & $\A$ & $\A$ & $\A$ & $\A$ & $\A$ & $\A$ & $\A$ & $\A$ &$\A$\\
$\overline {02}$ & $\overline {02}$ & $\overline {13}$ & $\overline {02}$ & $\overline {13}$ & $\A$ & $\overline {02}$ & $\A$ & $\A$ & $\overline {13}$ & $\A$ & $\A$ & $\A$ & $\A$ & $\A$ &$\A$\\
$\overline {03}$ & $\A$ & $\A$ & $\A$ & $\A$ & $\A$ & $\A$ & $\A$ & $\A$ & $\A$ & $\A$ & $\A$ & $\A$ & $\A$ & $\A$ &$\A$\\
$\overline {12}$ & $\A$ & $\A$ & $\A$ & $\A$ & $\A$ & $\A$ & $\A$ & $\A$ & $\A$ & $\A$ & $\A$ & $\A$ & $\A$ & $\A$ &$\A$\\
$\overline {13}$ & $\overline {13}$ & $\overline {02}$ & $\overline {13}$ & $\overline {02}$ & $\A$ & $\overline {13}$ & $\A$ & $\A$ & $\overline {02}$ & $\A$ & $\A$ & $\A$ & $\A$ & $\A$ &$\A$\\
$\overline {23}$ & $\A$ & $\A$ & $\A$ & $\A$ & $\A$ & $\A$ & $\A$ & $\A$ & $\A$ & $\A$ & $\A$ & $\A$ & $\A$ & $\A$ &$\A$\\
$\overline {012}$ & $\A$ & $\A$ & $\A$ & $\A$ & $\A$ & $\A$ & $\A$ & $\A$ & $\A$ & $\A$ & $\A$ & $\A$ & $\A$ & $\A$ &$\A$\\
$\overline {013}$ & $\A$ & $\A$ & $\A$ & $\A$ & $\A$ & $\A$ & $\A$ & $\A$ & $\A$ & $\A$ & $\A$ & $\A$ & $\A$ & $\A$ &$\A$\\
$\overline {023}$ & $\A$ & $\A$ & $\A$ & $\A$ & $\A$ & $\A$ & $\A$ & $\A$ & $\A$ & $\A$ & $\A$ & $\A$ & $\A$ & $\A$ &$\A$\\
$\overline {123}$ & $\A$ & $\A$ & $\A$ & $\A$ & $\A$ & $\A$ & $\A$ & $\A$ & $\A$ & $\A$ & $\A$ & $\A$ & $\A$ & $\A$ &$\A$\\
$\A$ & $\A$ & $\A$ & $\A$ & $\A$ & $\A$ & $\A$ & $\A$ & $\A$ & $\A$ & $\A$ & $\A$ & $\A$ & $\A$ & $\A$ & $ \A$ \\ \hline
\end{tabular}

\bigskip

Note that all considerations in this section are valid for the Clifford algebras $\cl(p,q)$ of all dimensions $n=p+q$. But we get meaningful results only for $n\geq4$, since for $n<4$ the concept of quaternion type coincides with the concept of rank $$\st{\overline k}{U}=\st{k}{U},\quad k=0, 1, \ldots n,\quad n<4.$$
For $n=4$ we have: $$\st{\overline 0}{U}=\st{0}{U}+\st{4}{U},\quad \st{\overline 1}{U}=\st{1}{U},\quad \st{\overline 2}{U}=\st{2}{U},\quad \st{\overline 3}{U}=\st{3}{U}.$$

\section{Subalgebras in the form of linear combinations of elements of the given types}
The method of quaternion typification of Clifford algebra elements allow us to prove a number of new properties of Clifford algebras.

In this section we denote $\cl_{\overline{k}}^\R(p,q)$ by $\,\overline{\textbf{k}}\,$ and $\cl_{\overline{k}}^\C(p,q)$ by $\,\overline{\textbf{k}}\oplus i\overline{\textbf{k}}\,$.

\begin{theorem}2.
a) The subspace
\begin{equation}
\overline{\textbf{02}}=\cl^\R_{even}(p,q)
\end{equation}
forms subalgebra of the real Clifford algebra $\cl^\R(p,q)$.\\
b) Subspaces
\begin{align}
\overline{\textbf{02}}&=\cl^\R_{even}(p,q),& \overline{\textbf{02}}\oplus i\overline{\textbf{02}}&=\cl^\C_{even}(p,q) ,\\
\overline{\textbf{02}}\oplus i\overline{\textbf{13}}&=\cl^\R_{even}(p,q)\oplus i\cl^\R_{odd}(p,q),& \overline{\textbf{0123}}&=\cl^\R(p,q)\nonumber
\end{align}
form subalgebras of the complex Clifford algebra $\cl^\C(p,q)$.\\

\end{theorem}

\proof. \, With the aid of written out above table the proof of this theorem is straightforward. $\blacksquare$

\begin{theorem}3.
a) Subspaces
\begin{equation}
\overline{\textbf{2}},\qquad\overline{\textbf{02}},\qquad \overline{\textbf{12}},\qquad \overline{\textbf{23}}
\end{equation}
of the real Clifford algebra $\cl^\R(p,q)$ are closed with respect
to the commutator $\quad U, V \rightarrow [U,V]$ and, hence, form Lie algebras w.r.t. the commutator.\\
b) Subspaces
\begin{eqnarray}
&&\overline{\textbf{2}},\qquad\overline{\textbf{02}},\qquad\overline{\textbf{12}},\qquad\overline{\textbf{23}},\qquad\overline{\textbf{0123}},\nonumber\\
&&\overline{\textbf{02}}\oplus i\overline{\textbf{02}},\qquad\overline{\textbf{12}}\oplus i\overline{\textbf{12}},\qquad\overline{\textbf{23}}\oplus i\overline{\textbf{23}},\label{subalg}\\
&&\overline{\textbf{2}}\oplus i\overline{\textbf{0}},\qquad\overline{\textbf{2}}\oplus i\overline{\textbf{1}},\qquad\overline{\textbf{2}}\oplus i\overline{\textbf{2}},\qquad\overline{\textbf{2}}\oplus i\overline{\textbf{3}},\nonumber\\
&&\overline{\textbf{02}}\oplus i\overline{\textbf{13}},\qquad\overline{\textbf{12}}\oplus i\overline{\textbf{03}},\qquad\overline{\textbf{23}}\oplus i\overline{\textbf{01}}\nonumber
\end{eqnarray}
of the complex Clifford algebra $\cl^\C(p,q)$ are closed with respect
to the commutator $\quad U, V \rightarrow [U,V]$ and, hence, form Lie algebras w.r.t the commutator.\\

\end{theorem}

\begin{theorem}4.
a) Subspaces
\begin{equation}
\overline{\textbf{0}},\qquad\overline{\textbf{01}},\qquad \overline{\textbf{02}},\qquad \overline{\textbf{03}}
\end{equation}
of the real Clifford algebra $\cl^\R(p,q)$ are closed with respect to the operation $\quad U, V \rightarrow \{U,V\}$ and form subalgebras of the Clifford algebra considered with respect to the operation $\quad U, V \rightarrow \{U,V\}$.\\
b) Subspaces
\begin{eqnarray}
&&\overline{\textbf{0}},\qquad\overline{\textbf{01}},\qquad\overline{\textbf{02}},\qquad\overline{\textbf{03}},\qquad\overline{\textbf{0123}},\nonumber\\
&&\overline{\textbf{01}}\oplus i\overline{\textbf{01}},\qquad\overline{\textbf{02}}\oplus i\overline{\textbf{02}},\qquad\overline{\textbf{03}}\oplus i\overline{\textbf{03}},\label{subalg2}\\
&&\overline{\textbf{0}}\oplus i\overline{\textbf{0}},\qquad\overline{\textbf{0}}\oplus i\overline{\textbf{1}},\qquad\overline{\textbf{0}}\oplus i\overline{\textbf{2}},\qquad\overline{\textbf{0}}\oplus i\overline{\textbf{3}},\nonumber\\
&&\overline{\textbf{01}}\oplus i\overline{\textbf{23}},\qquad\overline{\textbf{02}}\oplus i\overline{\textbf{13}},\qquad\overline{\textbf{03}}\oplus i\overline{\textbf{12}}\nonumber
\end{eqnarray}
of the complex Clifford algebra $\cl^\C(p,q)$ are closed with
respect to the anticommutator $\quad U, V \rightarrow \{U,V\}$ and
form subalgebras of the Clifford algebra considered with respect
to the operation $\quad U, V \rightarrow \{U,V\}$.

\end{theorem}

\proof. \, With the aid of (\ref{1}),(\ref{2}) (or see above tables) the proof of this theorem is straightforward.$\blacksquare$

\bigskip
Now we consider the notions of the pseudo-unitary group $\Wcl^\C(p,q)$ of the complex Clifford algebra and the Lie algebra $\wcl^\C(p,q)$ of the Lie group $\Wcl^\C(p,q)$ (see in \cite{Shirokov}).

Consider the following set of Clifford algebra elements:
\begin{equation}
\Wcl^\C(p,q)=\{U\in\cl^\C(p,q): U^* U=e\}, \label{psungr}
\end{equation}
where * is the operation of Clifford conjugation
\cite{Marchuk:Shirokov} with properties
$$
e^*=e,\quad (e^a)^*=e^a,\quad (\lambda\ e^{a_1}e^{a_2}\ldots e^{a_k})^*=\overline{\lambda}\ e^{a_k}\ldots e^{a_1},\quad
$$
$\lambda$ is a complex number and $\overline{\lambda}$ is the
conjugated complex number. This set forms a (Lie) group with respect
to the Clifford product and this group is called {\it the
pseudo-unitary group of the Clifford algebra $\cl(p,q)$ }.

The set of elements with the commutator $[U,V]=UV-VU$
\begin{equation}
\wcl^\C(p,q)=\{u\in\cl^\C(p,q): u^*=-u\}.\label{liealg}
\end{equation}
is {\it the Lie algebra of the Lie group $\Wcl^\C(p,q)$}.

From this definition and from the definition of Clifford conjugation  it follows that an arbitrary element of this Lie algebra has the form
$$
u=i\st{0}{u}+i\st{1}{u}+\st{2}{u}+\st{3}{u}+i\st{4}{u}+i\st{5}{u}+\ldots+
a_n\st{n}{u}=\sum_{k=0}^{n}a_k \st{k}{u},
$$
where $\st{k}{u} \in\cl_k^\R (p,q)$ and
$$
a_k=\left\lbrace
\begin{array}{ll}
1, & \mbox{\rm $k=2, 3, 6, 7, \ldots$;}\\
i, & \mbox{\rm $k=0, 1, 4, 5, \ldots$}
\end{array}
\right.
$$

So
\begin{equation}
\wcl^\C(p,q)=i\cl_{\overline{0}}^\R(p,q) \oplus i\cl_{\overline{1}}^\R(p,q) \oplus \cl_{\overline{2}}^\R(p,q) \oplus \cl_{\overline 3}^\R(p,q).\label{wcl}
\end{equation}

\begin{theorem}5. The Lie algebra $\wcl^\C(p,q)$ of the Lie group $\Wcl^\C(p,q)$ is an algebra of quaternion type with respect to the operation $\quad U, V \rightarrow [U,V]$ and $$\quad\bbE=\cl_{\overline{2}}^\R(p,q),\quad\bbI=\cl_{\overline{3}}^\R(p,q),\quad\bbJ=i\cl_{\overline{0}}^\R(p,q),\quad\bbK=i\cl_{\overline 1}^\R(p,q)\quad.$$

\end{theorem}

\proof. \, The statement of the theorem is equivalent to the following properties:
\begin{align}
[i\overline k,i\overline k]&\subseteq\overline{\textbf{2}},& k&=0, 1, \nonumber\\
[\overline k,\overline k]&\subseteq\overline{\textbf{2}},& k&=2, 3, \nonumber\\
[i\overline k,\overline 2]&\subseteq i\overline{\textbf{k}},& k&=0, 1, \label{th4}\\
[\overline k,\overline 2]&\subseteq\overline{\textbf{k}},& k&=3, \nonumber \\
[i\overline 0,i\overline 1]&\subseteq\overline{\textbf{3}},\quad [i\overline 0,\overline 3]\subseteq i\overline{\textbf{1}},\quad[i\overline 1,\overline 3]\subseteq i\overline{\textbf{0}} \nonumber.
\end{align}
But these formulas follow from (\ref{1}). These completes the proof of the theorem. $\blacksquare$

\begin{theorem}6.
Subspaces
\begin{equation}
\overline{\textbf{2}},\qquad\overline{\textbf{2}}\oplus i\overline{\textbf{0}},\qquad\overline{\textbf{2}}\oplus i\overline{\textbf{1}},\qquad\overline{\textbf{23}}
\end{equation}
of the complex Clifford algebra $\cl^\C(p,q)$ are closed with respect to the operation $\quad U, V \rightarrow [U,V]$ and form subalgebras of the Lie algebra $\wcl^\C(p,q)$ of the pseudo-unitary group of the Clifford algebra.\\

\end{theorem}

\proof. \, With the aid of (\ref{subalg}) and (\ref{wcl}) the proof of this theorem is straightforward.$\blacksquare$

\begin{theorem}7.
The following subspaces form subgroups of pseudo-unitary group $\Wcl^\C(p,q)$. The Lie algebras from Theorem 6 correspond to these Lie groups.\\
\bigskip
$$\begin{tabular}{|c|c|}\hline
Lie algebra & Lie group\\\hline
$\qquad\overline{\textbf{2}}\qquad$&$\{U\in\overline{\textbf{02}}=\cl^\R_{even}(p,q): U^* U=e\}$\\\hline
$\qquad\overline{\textbf{2}}\oplus i\overline{\textbf{0}}\qquad$&$\{U\in\overline{\textbf{02}}\oplus i\overline{\textbf{02}}=\cl^\C_{even}(p,q): U^* U=e\}$\\\hline
$\qquad\overline{\textbf{2}}\oplus i\overline{\textbf{1}}\qquad$&$\{U\in\overline{\textbf{02}}\oplus i\overline{\textbf{13}}=\cl^\R_{even}(p,q)\oplus i\cl^\R_{odd}(p,q): U^* U=e\}$\\\hline
$\qquad\overline{\textbf{23}}\qquad$&$\{U\in\overline{\textbf{0123}}=\cl^\R(p,q): U^* U=e\}$\\\hline
\end{tabular}$$
\bigskip

\end{theorem}

\proof. \, Let's prove, for example, the first of four statements. Let $U$ be an element of Lie group $\{U\in\overline{\textbf{02}}: U^* U=e\}$. Then
\begin{equation}
U=e+\varepsilon u,
\end{equation}
where $\varepsilon^2=0$ and $u$ - an element of the real Lie algebra of this Lie group (there is only one such Lie algebra). Then
\begin{equation}
e=U^*U=(e+\varepsilon u^*)(e+\varepsilon u)=e+\varepsilon(u+u^*).\nonumber
\end{equation}
So, for element of Lie algebra we have $u^*=-u$, i.e. $u\in\overline{\textbf{23}}\oplus i\overline{\textbf{01}}$. But also $u\in\overline{\textbf{02}}$. Thus, $u\in\overline{\textbf{2}}$.$\blacksquare$

\section{Conclusion}

In this paper we present a new classification of Clifford algebra elements, based on the notion of quaternion type. In many cases this classification is more suitable than the classification of Clifford algebra elements based on the notion of rank (\ref{ranks}) or classification based on the notion of parity (\ref{evenness}). New classification allows us to prove new properties and generalize results which are true only for small dimensions of Clifford algebras. We present these results in the following papers (\cite{Shirokov}, \cite{Sh1}, \cite{Sh2}).

The following Lie groups and Lie algebras - unitary, orthogonal, pseudo-orthogonal, symplectic, spinor - are widely used in mathematical and theoretical physics. The method of quaternion typification is used in the analysis of these groups and algebras. For example, in \cite{Shirokov} we give a classification of subalgebras of Lie algebras of pseudo-unitary groups on the basis of the method of quaternion typification.

\subsection*{Acknowledgment}
The author is grateful to
N.~G.~Marchuk for the constant attention to this work.

\end{document}